# Excitation of wakefields in a plasma channel by a laser pulse

## [1]R.Annou and V.K. Tripathi

Physics Department, Indian Institute of Technology, New Delhi-110 016, INDIA
[1] Permanent address: Faculty of physics, University of Sciences and Technology Houari Boumedienne, Algiers - Algeria

**Abstract**

The excitation of wakefields by a triangular temporal profile laser pulse with an abrupt fall in a step density profile plasma channel is investigated analytically and an exact solution is obtained. The excitation is more efficient for $\tau \sim \tau_p$, where $\tau$ is the rise time of the pulse and $\tau_p$ is the plasma period, though it is still significant for $\tau \sim 20\,\tau_p$. The wake potential in the former case is 1.7 times the ponderomotive potential, whereas it falls to 1.4 times the ponderomotive potential for the latter.

## I. Introduction

In linear accelerators, the accelerating gradients are severely limited due to the breakdown. Hence, the necessity to use an already ionized medium capable of sustaining very high electric fields, has been felt. Many schemes have been considered to excite large amplitude plasma waves, viz., the beat-wave, the plasma wake-field and the laser wake-field accelerators, to achieve high acceleration gradients [1,2]. However, the energy gain of the accelerated charged particle is reduced due to diffraction, for the plasma acts as a dielectric medium. To increase the acceleration energy gain, guiding of the laser pulse is needed[3].

In this note, we address the issue of excitation of plasma wake-fields by a laser pulse propagating in a step density profile plasma channel.

## II. Formulation

We consider a non evolving laser pulse of frequency $\omega_0$ with a triangular shape in time having an abrupt fall, propagating in a plasma channel. The evolution of the cold and collision less plasma is governed by the linearized continuity and momentum equations along with Poisson's equation; i.e.

$$\frac{\partial v_0}{\partial t} = \frac{e}{m} \nabla (\Phi_0 + \Phi_p)$$
$$\frac{\partial n_0}{\partial t} + \nabla \cdot n_{00} v_0 = 0 \qquad (1)$$
$$\nabla^2 \Phi_0 = 4\pi e n_0$$

The background density $n_{00}=n_{001}$ for r<a and $n_{002}$ otherwise, and the ponderomotive potential $\Phi_p = -\Phi_{p0} y(r) \cdot g(x)$, where $x = t - z/v_g$, $v_g$ is the group velocity of the laser pulse, y(r) is the laser pulse profile and $\Phi_{p0} = \frac{1}{2}\frac{m}{e}v_{osc}^2$, $v_{osc}$ being the quiver velocity.

After some algebra, the equation governing the plasma wave electrostatic potential $\Phi_0$ is obtained from Eqs. (1),

$$\nabla^2 (\frac{\partial^2}{\partial t^2}\Phi_0 + \omega_p^2 \Phi_0 + \omega_p^2 \Phi_p) = 0 \quad , \qquad (2)$$

where $\omega_p^2 = 4\pi n_{00} e^2 / m$

We use the Laplace transform in Eqs(2) to get,

$$\nabla^2 \left[ \tilde{\Phi}_0 - \Phi_{p0} y(r) \tilde{g}(x) \frac{w_p^2}{P^2 + w_p^2} \right] = 0 ,  \qquad (3)$$

where, $\tilde{f}(P) = \int_0^\infty e^{-Pt} f(t) dt$ is the Laplace transform of f(t).

The solution of Eq.(3) is,

$$\tilde{\Phi}_{01} = C_1 I_0(kr) e^{-ikz} + \Phi_{p0} y(r) \tilde{g}(x) \frac{w_{p1}^2}{P^2 + w_{p1}^2} , \quad r \leq a$$

$$\tilde{\Phi}_{02} = C_2 K_0(kr) e^{-ikz} + \Phi_{p0} y(r) \tilde{g}(x) \frac{w_{p2}^2}{P^2 + w_{p2}^2} , \quad r > a$$

(4)

where, $I_0$ and $K_0$ are Hankel functions. The constants $C_1$ and $C_2$ are obtained by virtue of the continuity of $\tilde{\Phi}_0$ and $(1 - w_p^2 / w^2) \frac{\partial \tilde{\Phi}_0}{\partial r}$ at r=a.

Considering $g(x) = (x/t)[u(x) - u(x-t)]$, where u(ξ) is the Heaviside function, one gets,

$$C_2 = C_1 I_0(ka) + \Phi_{p0} y(a) b(P) \left[ \frac{w_{p2}^2}{P^2 + w_{p2}^2} - \frac{w_{p1}^2}{P^2 + w_{p1}^2} \right] ,$$

$$C_1 = \frac{\Phi_{p0} b(P)(w_{p2}^2 - w_{p1}^2) \left[ \frac{y'(a)}{P^2} K_0(ka) + k K_1(ka) \frac{y(a)}{P^2 + w_{p1}^2} \right]}{k \left[ K_1(ka) I_0(ka)(1 + \frac{w_{p2}^2}{P^2}) - (1 + \frac{w_{p1}^2}{P^2}) I_1(ka) K_0(ka) \right]} \qquad (5)$$

with $k = \frac{P}{iv_g}$,

where, $b(P) = \frac{1}{tP^2}(e^{-tP}(1 + tP) - 1)$. The third relation stands for the equality of the phase velocity of the plasma wave and the group velocity of the pulse. Hence, the wake electrostatic potential $\Phi_0$ inside the channel may be obtained by inversing the Laplace transform,

$$\Phi_{01} = \frac{1}{2\pi} \int_{g-i\infty}^{g+i\infty} e^{xP} \left[ C_1(P) I_0\left(\frac{Pr}{iv_g}\right) + \Phi_{P0} y(r) b(P) \frac{w_{p1}^2}{P^2 + w_{p1}^2} \right] dP \tag{6}$$

It might be noted also that when the laser pulse is not taken into account, the channel eigen modes (or the modes of the plasma fiber) are simply,

$$\frac{w_{p2}^2 - w^2}{w_{p1}^2 - w^2} = -\frac{K_0(ka) I_1(ka)}{K_1(ka) I_0(ka)} \tag{7}$$

In Fig.[1], the linear dispersion relation $w = w(k)$ is plotted. It is clear that the frequency of the mode is within a range defined by the plasma frequencies $w_{p1}^2 < w < w_{p2}^2$.

The integral in Eq.(6) possesses residues at P=0, $iw_{p1}$, and $iw_s$ where ω$_s$ represents the wave satisfying the dispersion relation given by Eq.(6), and the third relation of Eq.(5). However, the pole P=0 is not considered since we are seeking oscillating solutions.

The solution of Eq.(6) when the pulse is gone is,

$$\frac{\Phi_{01}}{\Phi_{p0}} = -\frac{\sin(w_{p1}x)}{\tau w_{p1}} \left[ \frac{y(a) I_0\left(\frac{rw_{P1}}{v_g}\right)}{2 I_0\left(\frac{aw_{p1}}{v_g}\right)} + y(r) \right] +$$

$$+ \left( \cos(w_{p1}(x-\tau)) + \frac{\sin(w_{p1}(x-\tau))}{\tau w_{p1}} \right) \left[ \frac{y(a) I_0\left(\frac{rw_{p1}}{v_g}\right)}{2 I_0\left(\frac{aw_{p1}}{v_g}\right)} + y(r) \right] - \tag{8}$$

$$-y \left( \cos(w_s(x-\tau)) + \frac{\sin(w_s(x-\tau))}{\tau w_s} - \frac{\sin(w_s x)}{\tau w_s} \right)$$

where,

$$y = I_0\left(\frac{rw_s}{v_g}\right) \frac{y(a) w_s K_1\left(\frac{aw_s}{v_g}\right) + y'(a)\left(\frac{v_g}{w_s^2}\right)\left(w_s^2 - w_{p1}^2\right) K_0\left(\frac{aw_s}{v_g}\right)}{2 I_0\left(\frac{aw_s}{v_g}\right) w_s + \frac{a}{v_g}\left(w_s^2 - w_{p1}^1\right)\left( I_0\left(\frac{aw_s}{v_g}\right) K_0\left(\frac{aw_s}{v_g}\right) - I_1\left(\frac{aw_s}{v_g}\right) K_1\left(\frac{aw_s}{v_g}\right) \right)}.$$

## III. Discussion

After the pulse is gone, the Langmuir wave is produced and its potential is brought to an amplitude that depends on $\tau$. The value of $\tau$ yielding maximum value of $\Phi_0$ at the instant the pulse is gone, is $\tau\omega_{p1} \sim 4.5$. Hence, if we consider a channel depth of $(n_{002}-n_{001})/n_{002} \sim 50\%$ (see Ref.(4)), we obtain $\tau \sim \tau_p$ where $\tau_p \sim 2\pi/\omega_{p2}$. It is to be pointed out however, that for $\tau \sim 20\tau_p$ we still have a wave excited with an amplitude at the moment the pulse is gone given by $\Phi_{01}/\Phi_{p0} \sim 4/3$. The maximum electron energy gain per unit length is of the order of 8GeV/m. For typical values of the system, i.e., $n_{002} \sim 10^{16}$ cm$^{-3}$, $v_g \sim c$, $a=75\mu$m, $a/r_0 \sim 0.7$, $\omega_s \sim 1.3\omega_{p1}$, where y (r)=exp(-r$^2$/r$_0^2$), we have plotted $\Phi_{01}/\Phi_{p0}$ vs. $\xi/\tau$ on the axis (r=0) in Fig.[2]. On the other hand $\Phi_{01}/\Phi_{p0}$ vs r/a for $\xi=\tau$, has been plotted in Fig.[3].

The excited wave has a potential $\Phi_0$ linearly dependent on and has the same order of magnitude as the ponderomotive potential $\Phi_{p0}$. The plasma wave excitation is the most efficient for $\tau \sim \tau_p$. However, for long pulses, a sensitive plasma wave generation is expected as well. It may be noted also that the amplitude of the wave is modulated at a frequency $\sim 0.17\omega_{p1}$

### Acknowledgement
This work has been supported by DST (INDIA,1998) and USTHB (ALGERIA).

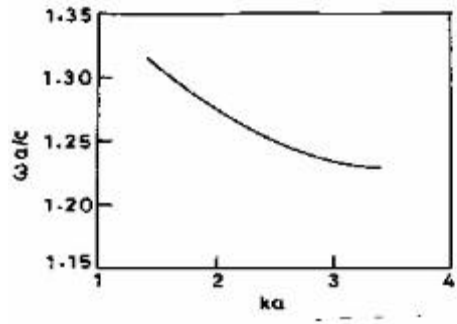

Fig.[1]. Linear dispersion relation $w = w(k)$, of the channel eigen modes.

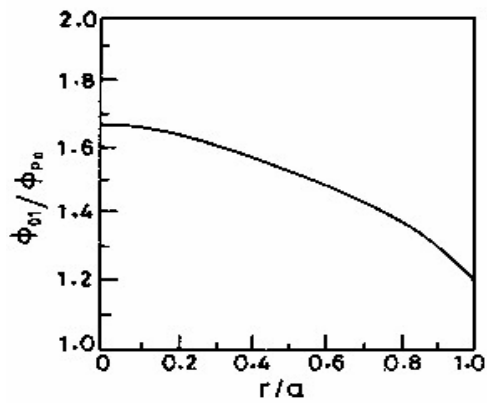

Fig.[2]. Evolution of normalized wake potential on the axis (r=0) with respect to time.

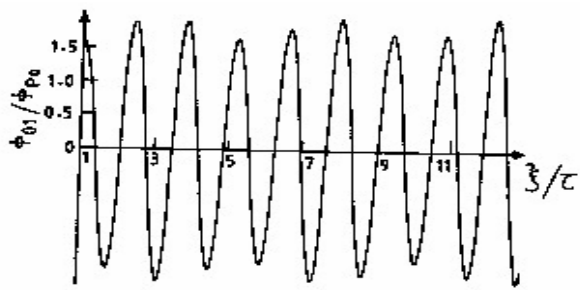

Fig.[3]. Evolution of normalized wake potential at $x = t$, with respect to the normalized radius (r/a) in the channel.